\documentclass[pra,aps,a4paper,twocolumn,repreprint,longbibliography]{revtex4}

\usepackage{amssymb}
\usepackage{amsmath}
\usepackage{amsfonts}
\usepackage{graphicx}
\usepackage{bm}
\usepackage[dvipsnames]{xcolor}
\usepackage{multirow}
\usepackage{natbib}
\usepackage{hyperref}
\usepackage{relsize}

\DeclareMathOperator{\erf}{erf}

\newcommand\tit[1]{\emph{#1},}

\newcommand{\eps}{\varepsilon}
\newcommand{\corr}[1]{\langle #1\rangle}

\def\be{\begin{equation}}
\def\ee{\end{equation}}

\begin{document}

\title{Lyapunov exponent for Whitney's problem with random drive}

\author{Nikolai A.~Stepanov}
\affiliation{Skolkovo Institute of Science and Technology, Moscow 121205, Russia}
\affiliation{L.~D.~Landau Institute for Theoretical Physics, Chernogolovka 142432, Russia}

\author{Mikhail A.~Skvortsov}
\affiliation{Skolkovo Institute of Science and Technology, Moscow 121205, Russia}
\affiliation{L.~D.~Landau Institute for Theoretical Physics, Chernogolovka 142432, Russia}



\begin{abstract}
We consider the statistical properties of a non-falling trajectory in the
Whitney problem of an inverted pendulum excited by an external force. In the
case when the external force is white noise, we recently found the
instantaneous distribution function of the pendulum angle and velocity over
an infinite time interval using a transfer-matrix analysis of the
supersymmetric field theory. Here, we generalize our approach to the case of
finite time intervals and multipoint correlation functions. Using the
developed formalism, we calculate the Lyapunov exponent, which determines
the decay rate of correlations on a non-falling trajectory.
\end{abstract}

\date{August 27, 2020}
\maketitle

\textbf{1.}
Balancing an inverted pendulum under a given time-dependent horizontal force
$f(t)$ is a famous mathematical problem formulated by Courant and Robbins in
their book \emph{What is Mathematics?} (first edition in 1941) \cite{CR-book},
where Whitney was credited as the author of the problem. Using fairly
general mathematical arguments based on the intermediate value theorem, they showed
that for any force $f(t)$ acting during a finite time interval $[0,T]$, an
initial position of the pendulum in the upper half-plane can be chosen such
that it will remain in the upper half-plane during the further evolution for
all $t\in[0,T]$. The existence of a \emph{non-falling trajectory} (non-FT) in
the Whitney problem has been the subject of an ongoing debate in the
mathematical literature \cite{Broman,Poston}, resulting in a critical
analysis and refinement of the original arguments of Courant and Robbins.
Fresh interest in the problem of an inverted pendulum is associated with
Arnold, whose view in 2002 was that this problem still awaits a rigorous
solution \cite{Arnold}. In 2014, Polekhin presented a proof of the existence
of the non-FT using the Wazewski topological principle \cite{W-proof}. This work
provoked several publications generalizing his approach and proposing new
topological methods \cite {Zubelevich,BolKoz,Srzednicki} (see Refs.\
\cite{Srzednicki,Shen} for good reviews of the history of the Whitney
problem).

\begin{figure}[t!]
\includegraphics[width=0.97\columnwidth]{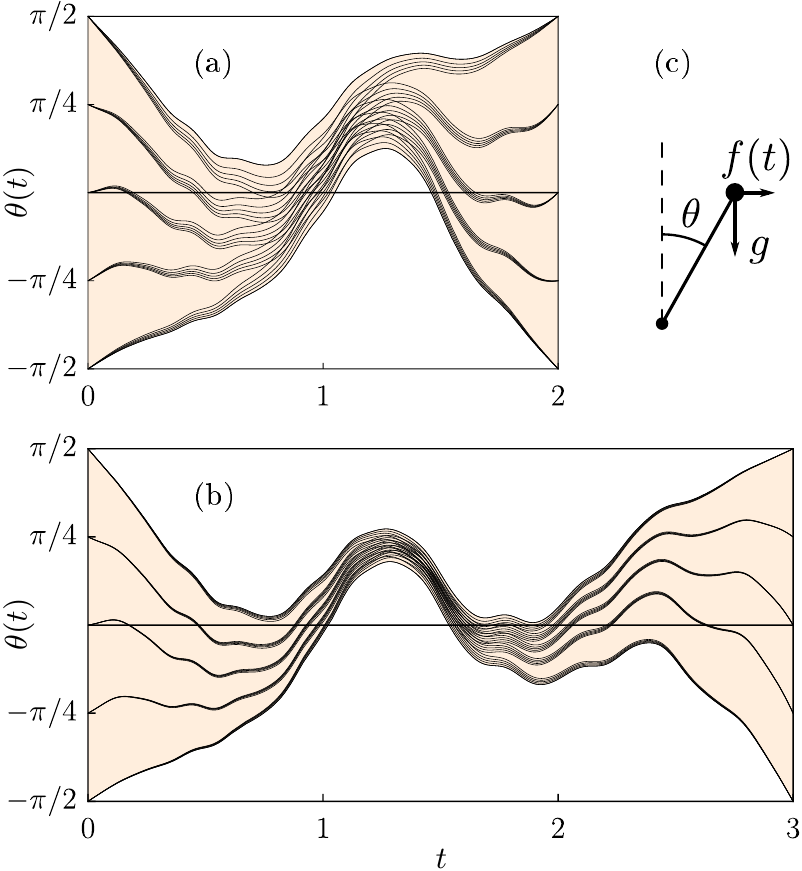}
\caption{Examples of non-falling trajectories for the pendulum equation of motion
\eqref{eq} considered on two time intervals: (a)~$T=2/\omega$
and (b) $T=3/\omega$. For any choice of $\theta_1$ and $\theta_2$ in the upper
half-plane ($|\theta|<\pi/2$), there exists a unique non-falling solution
satisfying the boundary conditions $\theta(0)=\theta_1$ and $\theta(T)=\theta_2$.
In the plots, we show 25 such trajectories with $\theta_{1,2}=(-1,-0.5,0,0.5,1)\times\pi/2$.
The driving force is $f(t)=4\sum_{n=1}^{40}\cos(kt+k^4)$ in both cases. (c)~An
inverted pendulum under the action of a horizontal force.
}
\label{F:veers}
\end{figure}

Recently we developed a theory of the statistical description of a
\emph{never falling trajectory} (NFT) of an inverted pendulum under the
action of a random force~\cite{we}. An NFT can be regarded as the limit of
non-FTs in the Whitney problem as the length $T$ of the time interval tends
to infinity. The NFT concept is illustrated in Fig.~\ref{F:veers}, which shows 
numerical solutions to the boundary value problem for the pendulum equation
(the angle $\theta$ is measured from the vertical)
\be
\label{eq}
  \ddot\theta = \omega^2 \sin\theta + f(t) \cos\theta
\ee
with different initial and final values $\theta(0)=\theta_1$ and
$\theta(T)=\theta_2$ and a sufficiently rapidly varying force $f(t)$. For
any $\theta_{1,2}$ in the strip $-\pi/2<\theta_{1,2}<\pi/2$, a non-falling
solution ($-\pi/2<\theta(T)<\pi/2$) of this boundary value problem exists
and is unique~\cite{we}. As $\theta_1$ and $\theta_2$ run through all
possible values in the strip, the set of corresponding non-FTs form a
bundle, shown in color in Fig.~\ref{F:veers}. This bundle shrinks as one
moves away from the boundary, becoming exponentially thin in the middle of
the interval for large $T$. In the limit $T\to\infty $, when the pendulum
must be balanced on the entire real axis, the non-FT bundle for the Whitney
problem on a finite time interval becomes infinitely thin and defines a
unique \emph{never falling trajectory}, which is a functional of the given force $f(t)$.

In Ref.\ \cite{we}, we studied the statical properties of an NFT in the case
when the driving force is Gaussian white noise with the correlator
\be
\label{corr}
  \corr{f(t)f(t')} = 2 \alpha \delta(t-t'),
\ee
and calculated the instantaneous distribution function $P(\theta,p)$ of
the angle $\theta$ and its velocity $p=\dot\theta$. Our approach is based on
the supersymmetric field theory formulation of stochastic dynamics proposed
by Parisi and Sourlas~\cite{PS1,PS2,Zinn}, which allows averaging over the
random force at the very beginning of the calculations. It is essential that
for the considered problem, the Parisi--Sourlas method is free from the
problem of the sign of the fermionic determinant because of the uniqueness of the non-FT. 
Using the idea of reducing the one-dimensional functional integral
to an effective quantum mechanics~\cite{EL83}, we were able to express the
distribution function $P(\theta,p)$ in terms of the zero mode of the
transfer-matrix Hamiltonian, which reduces to the Fokker--Planck operator
with a special type of boundary conditions ensuring that the trajectories do
not leave the strip.

Here, we extend the ideas of Ref.\ \cite{we} and consider a range of issues
related to the Lyapunov exponent for a non-FT. The Lyapunov exponent
determines both the law of the convergence of a non-FT on a finite time
interval to the NFT on an infinite time interval (see Fig.~\ref{F:veers})
and the decay of different-time correlators on the NFT. From the
technical standpoint, our result consists in describing the entire spectrum
of the transfer-matrix Hamiltonian, whose zero mode was studied in Ref.\ \cite{we}. 
In this language, the Lyapunov exponent is determined by the energy
of the first excited state. The developed theory allows calculating any
correlation functions for a non-FT on infinite, semi-infinite, and finite
time intervals.

\begin{figure}
\includegraphics[width=\columnwidth]{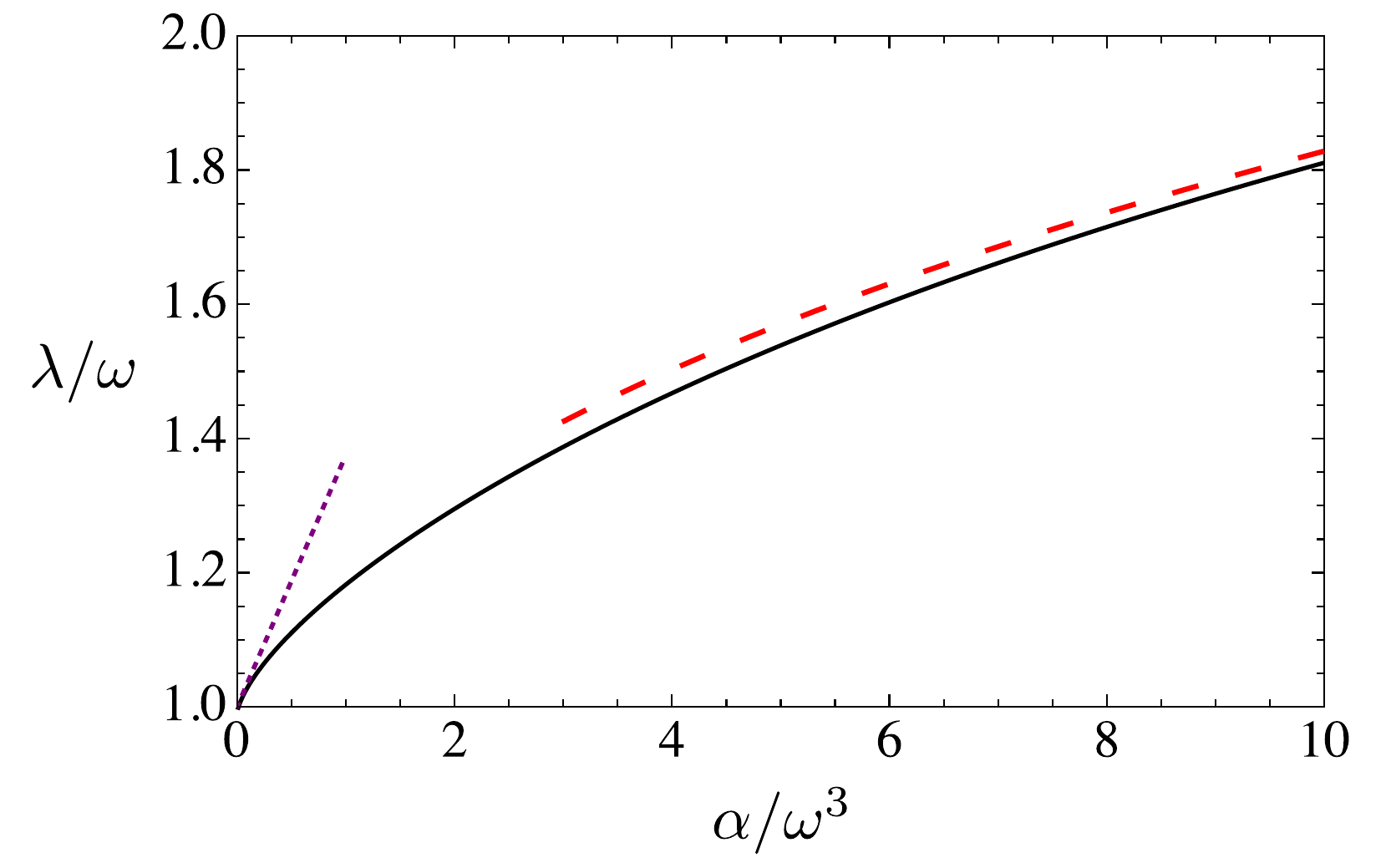}
\caption{Dependence of the Lyapunov exponent for a non-FT on the driving
strength measured by the parameter $\alpha/\omega^3$: the dotted line shows the
linear part of the asymptotic form for small $\alpha$, and the dashed line
shows the first three terms of expansion~\eqref{lambda-result} for large
$\alpha/\omega^3$.
}
\label{F:lambda}
\end{figure}

We show that the Lyapunov exponent in the Whitney problem with white-noise 
driving \eqref{corr} can be written as
\be
  \lambda = \omega g(\alpha/\omega^3) ,
\ee
where the function $g(x)$ has the asymptotic behavior
\be
\label{lambda-result}
  g(x) =
  \begin{cases}
  \displaystyle
    1
    + \frac{3}{8} x
    - \frac{525}{1024} x^2
    + \dots , & x\ll1 , \\[9pt]
     0.66  \, x^{1/3} + 0.26 + 0.30\, x^{-1/3}\dots, & x\gg1 .
  \end{cases}
\ee
In the absence of driving ($\alpha=0$), the Lyapunov exponent
$\lambda=\omega$ is determined by the exponential instability of the
trajectories near the upper pendulum position. For weak driving
($\alpha/\omega^3\ll1$), the typical non-FT angle is of the order
$\theta\sim(\alpha/\omega^3)^{1/2}$~\cite{we}, and the nonlinearity of
Eq.~\eqref{eq} leads to an increase in the Lyapunov exponent, which can be
expanded in an asymptotic series in powers of the small parameter
$\alpha/\omega^3$. Finally, under strong driving ($\alpha/\omega^3\gg1$),
the Lyapunov exponent becomes independent of $\omega$, reaching the limiting
value $\lambda\approx0.66\,\alpha^{1/3}$. We show the numerically found
dependence of the Lyapunov exponent on $\alpha/\omega^3$ in Fig.~\ref{F:lambda}.

\textbf{2.}
According to the approach developed in Ref.~\cite{we}, the statistical properties
of a non-FT are expressed in terms of the two-component ``wave function''
$\hat\Psi(\theta,p)=(\Psi,\Phi)^T $, whose evolution is governed by the imaginary-time
Schr\"odinger equation with the corresponding transfer-matrix Hamiltonian:
\be
\label{QM}
  \frac{\partial}{\partial t}
  \begin{pmatrix}
    \Psi \\
    \Phi
  \end{pmatrix}
  =
  -
  {\cal H}
  \begin{pmatrix}
    \Psi \\
    \Phi
  \end{pmatrix} ,
\qquad
  {\cal H}
  =
  \begin{pmatrix}
    L & {  -1 } \\
    { V_2 } & L
  \end{pmatrix} ,
\ee
where $L$ is the Fokker--Planck operator for the Kramers
problem~\cite{Risken-FPE},
\be
\label{L-FP}
  L = p\partial_{\theta}
  + \omega^{2}\sin\theta \, \partial_{p}
  - \alpha\cos^{2} \theta \, \partial_{p}^{2} ,
\ee
and the potential $V_2$ has the form
\be
\label{V2}
  V_2
  =
- \omega^2 \cos\theta
 - \alpha \sin2\theta \, \partial_p .
\ee

In Ref.\ \cite{we}, we studied the one-point correlation function of the NFT on
the infinite time interval, which is determined by the zero mode
$\hat\Psi_0$ of the Hamiltonian $\cal H$. Finding the zero mode is
significantly simplified due to the presence of the Becchi-Rouet-
Stora-Tuytin (BRST) symmetry of the action in the Parisi--Sourlas
representation of stochastic dynamics~\cite{Zinn}, 
which allows expressing both components of
$\hat\Psi(\theta,p)$ in terms of a scalar superpotential $\psi(\theta,p)$
via
\be
\label{BRST-reduct}
  \Psi =  \partial_p \psi ,
\qquad
  \Phi = -  \partial_\theta \psi .
\ee
The time evolution of $\psi$ is determined by the Fokker--Planck
operator $L$:
\begin{equation}
\label{FP_eq}
  \frac{\partial\psi}{\partial t}
  =
  -
L \psi
.
\end{equation}
But reduction~\eqref{BRST-reduct} works neither for calculating multi-time
correlation functions of the NFT nor for describing the statistics of a
non-FT on bounded intervals. In the former case, the BRST symmetry is
broken by the operators of physically observable quantities acting
identically on the wave function components $\Psi$ and $\Phi$. In the latter
case, the BRST symmetry is broken by the BRST-asymmetric initial condition at
the boundary of the interval [see Eq.~\eqref{Psi-bnd} below]. In both cases,
to describe the non-FT statistics, one must work with the two-components 
wave function $(\Psi,\Phi)$ and understand the properties of the Hamiltonian
$\cal H$.

We start with discussing the initial condition for the wave function at the boundary 
of an interval. To ensure that the non-FT is unique, we must fix the value of
$\theta$ at the boundary. (Generally speaking, one can fix the value of
$\dot\theta$ or even a linear combination of $\theta$ and $\dot\theta$, but
for simplicity, we assume that the angle is given.) By construction, the
wave function $\hat\Psi$ is closely related to the partition function of the
supersymmetric functional integral~\cite{we}. Right at the boundary, it
cannot contain Grassmann variables, which leads to the component $\Phi$
vanishing. Hence, the wave function at the interval boundary with the fixed value
$\theta=\theta_0$ has the form
\be
\label{Psi-bnd}
  \hat\Psi_{\theta_0}^\text{(b)}
  =
  \begin{pmatrix}
    \delta(\theta-\theta_0) \\ 0
  \end{pmatrix}
.
\ee

Consider the boundary value problem $[T_\text{L},T_\text{R}]$ with the
boundary conditions $\theta(t_\text{L})=\theta_\text{L}$  and
$\theta(t_\text{R})=\theta_\text{R}$. The essence of the reduction of the
Parisi--Sourlas integral to the quantum mechanics \eqref{QM} is that the
correlation function $\corr{O_1(t_1)O_2(t_2)\dots}$ of physical quantities
$O_i$ at the instants $t_i$ ($t_1<t_2<\dots$) can be represented as the matrix
element
\be
\label{corr-observ}
  \corr{ \hat\Psi_{\theta_\text{L}}^\text{(b)} | \dots
    O_2(t_2) e^{-{\cal H}(t_2-t_1)}
    O_1(t_1) e^{-{\cal H}(t_1-t_\text{L})}
  | \hat\Psi_{\theta_\text{L}}^\text{(b)} } ,
\ee
where the scalar product of two wave functions is defined as~\cite{we}
\be
\label{scalar-prod}
  \corr{\hat\Psi | \hat\Psi'}
  =
  \int d\theta\, dp\,
  [
    \Psi(\theta,p)\Phi'(\theta,-p)
  + \Phi(\theta,p)\Psi'(\theta,-p)
  ]
  .
\ee
In Ref.\ \cite{we}, we studied the instantaneous joint distribution function
$P(\theta,p)$ of the angle and velocity on the NFT corresponding to the
operator $O=\delta(\theta-\theta_0)\delta(p-p_0)$. Replacing
$\hat\Psi_{L,R}$ with the zero mode and using Eq.~\eqref{BRST-reduct}, one
can express $P(\theta,p)$ in terms of the Poisson bracket of the
superpotential $\psi$:
\begin{equation}
\label{P-via-Psi}
  P(\theta,p)
  =
  \bigl\{
    \psi(\theta,p), \psi(\theta,-p)
  \bigr\}_{\theta,p} .
\end{equation}

Both the Hamiltonian in Eq.~\eqref{QM} and the Fokker--Planck operator~\eqref{L-FP} 
are non-Hermitian. Generally speaking, such operators can lack
a complete system of eigenfunctions. However it is known that in the presence 
of friction the Fokker--Planck operator can be diagonalized \cite{Risken-FPE}, 
which makes it possible to construct a system of biorthogonal eigenfunctions 
and work with them practically as with eigenfunctions of a Hermitian 
operator~\cite{Jordan}. But there is no friction in our case, 
and we should therefore expect that the operators $\cal H$ and
$L$ reduce to the Jordan normal form. This results not in a simple
exponential decay of correlators as $t\to\infty$ but in the appearance of
additional powers of time [e.g., as can be seen in
expression~\eqref{corr-lin-inf}].

\textbf{3.}
To illustrate the developed approach, we consider the case of a \emph{weak
noise} ($\alpha/\omega^3\ll1$) in detail, where the Jordan structure of the
operators $\cal H$ and $L$ can be studied analytically. We start with the
Fokker--Planck operator. In the considered limit, the deviation of the
pendulum from the vertical is small ($\theta\ll1$), and the operator \eqref{L-FP} 
can be replaced with 
\be
\label{L-FP-lin}
  L = p\partial_{\theta}
  + \omega^{2}\theta \partial_{p}
  - \alpha \partial_{p}^{2} .
\ee
The zero mode of this operator corresponding to the NFT has the form
\be
\label{psi0-lin}
  \psi_0(\theta,p)
  =
  \erf(z)/2 ,
\ee
where we introduce ``holomorphic'' and ``antiholomorphic'' coordinates with
different signs of the momentum,
\be
\label{zz-def}
  z = \kappa\,(p-\omega\theta)
, \quad
  \overline z = - \kappa\,(p+\omega\theta) ,
\ee
where $\kappa=\sqrt{{\omega}/{2\alpha}}$. The spectrum of the
operator~\eqref{L-FP-lin} can be found using the identity $[L,\partial_z]=
\omega\partial_z$, which allows generating the eigenfunctions by
consecutively differentiating the zero mode with respect to $z$. We thus
find the eigenfunction of the $n$th excited state ($n=1,2,3,\dots$) with the
energy $\epsilon_n=n\omega$:
\be
  \psi_n
  =
  \frac{1}{\sqrt{\pi}}
  H_{n-1}(z) e^{-z^2} ,
\ee
where $H_n(z)=(-1)^ne^{z^2}d^ne^{-z^2}/dz^n$ is the Hermite polynomial (in
the physical definition). But the functions $\psi_n(\theta,p)$ thus
constructed depend only on the difference $p-\omega\theta$ (do not
contain $\overline z$) and therefore do not form a complete basis. This
circumstance is related to the fact that the non-Hermitian operator~\eqref{L-FP}
can be brought to the Jordan normal form and in addition to the eigenfunction has
several generalized eigenfunction corresponding to the same eigennumber $\eps_n$. 
It is easy to verify that the eigenfunction $\psi_n$ has $n-1$ generalized 
eigenfunctions, which we choose in the form
\be
\label{basis}
  \psi_{n,k}
  =
  \frac{(-1)^k}{2^kk!\sqrt\pi}
  H_{k}(\overline z) H_{n-k-1}(z) e^{-z^2} ,
\ee
where the index $k$ ranges from~1 to $n-1$. Together with $\psi_{n,0}=\psi_n$,
they form the basis of a Jordan block of dimension $n$ corresponding to the
energy $\epsilon_n=n\omega$:
\be
\label{J-block}
  L \psi_{n,k}
  =
  \epsilon_n \psi_{n,k} + \omega \psi_{n,k-1}
\ee
(to truncate the chain at the eigenfunction $\psi_{n,0}$, we set
$\psi_{n,-1}=0$).

The constructed system of functions is complete. An arbitrary function can
be decomposed with respect to the basis $\psi_{n,k}$ using the orthogonality
relation
\be
\label{ortho}
  \corr{\psi_{n,k}|\psi_{n',k'}}_z
  =
  (-1)^{n-1}
  \delta_{n,n'} \delta_{k+k'+1,n} ,
\ee
where the scalar product $\corr{\,\cdot\,|\,\cdot\,}_z$ is defined as
\be
\label{scalar-z}
  \corr{\psi|\psi'}_z
  =
  \int dz \, d\overline z \,
  \psi(\overline z,z)
  \psi'(z,\overline z) ,
\ee
and exchanging the arguments in one of the functions thus agrees with the
sign change for $p$ in Eq.~\eqref{scalar-prod}. We note that the integration
measures in Eqs.\ \eqref{scalar-prod} and~\eqref{scalar-z} are related by
$dz\,d\overline{z}=2\omega\kappa^2d\theta\,dp$.

During the evolution of the wave function $\psi_{n,k}$ under the action of
the operator $L$, other states of the Jordan block corresponding to the same
energy are mixed into it, which leads to the appearance of powers of $t$ on
top of the exponential decay:
\be
\label{exp-Lt}
  e^{-Lt}
  \psi_{n,k}
  =
  e^{-n \omega t}
  \sum_{m=0}^k
  \frac{(-\omega t)^m}{m!}
  \psi_{n,k-m} .
\ee

We now turn to studying the spectral properties of the Hamiltonian $\cal H$
in Eq.~\eqref{QM}. In the considered case of weak noise, Eq.~\eqref{V2}
gives $V_2=-\omega^2$, which partitions the state space of $\cal H$ into
even and odd sectors with the wave functions $\hat\Psi_{e,o}=(\Psi,\pm\omega\Psi)^T$
evolving independently with the Hamiltonians ${\cal H}_{e,o}=L\mp\omega$. The
system of eigenfunctions and generalized eigenfunctions of the operator $L$ 
constructed above thus allows completely describing the evolution of the doublet
$\hat\Psi$ under the action of the Hamiltonian $\cal H$.

Consider the evolution of the wave function~\eqref{Psi-bnd} 
away from the boundary in the limit $\alpha/\omega^3\ll1$.
Decomposing it into even and odd components, we obtain
\be
\label{Psi-bnd(t)}
  e^{-{\cal H}t} \hat\Psi_{\theta_0}^\text{(b)}
  =
  \begin{pmatrix}
    \cosh\omega t \\ \omega\sinh\omega t
  \end{pmatrix}
  e^{-Lt} \delta(\theta-\theta_0) .
\ee
To calculate the evolution of the delta function, we expand it in the basis
$\psi_{n,k}$:
\be
\label{delta}
  \delta(\theta-\theta_0)
  =
  \sum_{n=1}^\infty \sum_{k=0}^{n-1} c_{n,k} \psi_{n,k} .
\ee
The coefficients $c_{n,k}$ can be obtained using orthogonality
relations~\eqref{ortho} and the properties of Hermite polynomials
\be
  H_n(x+y)
  =
  \sum_{m=0}^n {n \choose k} (2y)^{n-k} H_k(x)
\ee
following from the Taylor expansion, and are given by
\be
\label{delta-expansion}
  c_{n,k}
  =
  (-1)^{n-1}
  2\kappa\omega
  \frac{(\theta_0/2\kappa\omega)^{n-2k-1}}{(n-2k-1)!} .
\ee
The evolution of the delta function in Eq.~\eqref{Psi-bnd(t)} follows from
expansion~\eqref{delta} and relations~\eqref{exp-Lt}. The memory of the
boundary is lost in the characteristic time $\omega^{-1}$ (the inverse
Lyapunov exponent). During this time, the difference between the two
components of the wave function $\hat\Psi$ is lost, and they both take the value
determined by the state $\psi_{1,0}$ with the minimum energy
$\epsilon_1=\omega$:
\be
\label{Psi-b(t)}
  \lim_{t\to\infty}
  e^{-Ht} \hat\Psi_{\theta_0}^\text{(b)}
  =
  \hat\Psi_0
  =
  \begin{pmatrix}
    1 \\ \omega
  \end{pmatrix}
  \kappa
  \psi_{1,0} ,
\ee
which is just the zero mode of \eqref{QM} in the limit
$\alpha/\omega^3\ll1$.

\textbf{4.}
We show how the developed spectral theory of the operators $\cal H$ and $L$
allows systematically calculating various correlation functions of the
non-FT \emph{in the case of weak noise}. The results in this section can
also be obtained directly by using the explicit expression for the non-FT in
terms of $f(t)$ with subsequent averaging over Gaussian white
noise~\eqref{corr} \cite{we}, but deriving them using the transfer-matrix
formalism is important methodologically because it illustrates the general
scheme and allows verifying its workability.

We begin by considering the calculation of the pair correlator for the NFT angle
on the entire real axis. Substituting the zero mode \eqref{Psi-b(t)}
into the general formula~\eqref{corr-observ} and taking into account that only
the even sector of the theory does contribute, we can express the 
correlator in terms of the scalar product~\eqref{scalar-z} in the
$z$-representation as
\be
\label{th-th-braket}
  \corr{\theta(0)\theta(t)}
  =
  \langle \psi_{1,0} | \theta e^{-(L-\omega)t} \theta | \psi_{1,0} \rangle_z .
\ee
Using Eqs.~\eqref{zz-def}  and \eqref{basis}, we can express
$\theta\psi_{1,0}$ in terms of the functions $\psi_{2,0}$ and $\psi_{2,1}$.
Then using the evolution law~\eqref{exp-Lt}, we obtain
\be
\label{th-psi-t}
  e^{-(L-\omega)t}
  \theta \psi_{1,0}
  =
  e^{-\omega t}
  \frac{\psi_{2,1}-(1/2+\omega t)\psi_{2,0}}{2\kappa\omega} .
\ee
Calculating the matrix element~\eqref{th-th-braket} as the overlap between the
states $e^{-(L-\omega)t}\theta\psi_{1,0}$  and $\theta\psi_{1,0}$ with the help of
relations~\eqref{ortho}, we find the sought pair correlator:
\be
\label{corr-lin-inf}
  \corr{\theta(0)\theta(t)}
  =
  \corr{\theta^2}
  (1+\omega t)
  e^{-\omega t}
 ,
\ee
where, as obtained in Ref.\ \cite{we},
\be
\label{th2-lin}
  \corr{\theta^2}=\alpha/2\omega^3 .
\ee
The appearance on the background $e^{-\omega t}$ of a contribution linearly
increasing with time is related to excitation of the states $\psi_{2,0}$ and
$\psi_{2,1}$ corresponding to the Jordan block of dimension~2.

In a similar way one can calculate more complicated correlators of the NFT. For
example,
\be
\label{corr22-lin-inf}
  \corr{\theta^2(0)\theta^2(t)}
  =
  \corr{\theta^2}^2
  \left[
  1+ 2(1+\omega t)^2 e^{-2\omega t}
  \right] .
\ee
Formally, the operator $\theta^2$ here applied to $\psi_{1,0}$ excites the
Jordan triplet $\psi_{3,0}$, $\psi_{3,1}$, $\psi_{3,2}$, which leads to the
appearance of terms up to $t^2$ on the background of the exponential
decay. But the structure of the correlator~\eqref{corr22-lin-inf} is related to the
Gaussian statistics of $\theta$ on the NFT~\cite{we}, which allows expressing
it in terms of pair correlator~\eqref{corr-lin-inf} using the Wick theorem.
Generalizing the developed formalism to multipoint correlators is also straightforward.

As the next example, we consider the calculation of the average angle
$\corr{\theta(t)}_{\theta_0}$ for the non-FT on the semi-infinite time
interval $t>0$ with the boundary condition $\theta(0)=\theta_0$. According
to Eq.\ \eqref{corr-observ}, the average angle is given by the matrix element
$\corr{\theta(t)}_{\theta_0}=\corr{\hat\Psi_0|\theta e^{-{\cal H}t}|
\hat\Psi_{\theta_0}^\text{(b)}}$. It is easiest to calculate by convoluting
expression~\eqref{th-psi-t} with the wave function~\eqref{Psi-bnd} at the
boundary. Integrating over the momentum, we see that the contribution from
$\psi_{2,0}=2ze^{-z^2}/\sqrt\pi$ disappears because it is odd in $z$, and we
obtain the simple exponential decay
\be
\label{corr-lin-semiinf}
  \corr{\theta(t)}_{\theta_0}
  =
  \theta_0
  e^{-\omega t} .
\ee
One can derive the same expression differently by calculating the matrix element
$\theta$ between the zero mode $\psi_{1,0}$ and evolved boundary wave
function~\eqref{Psi-bnd(t)}. Such matrix elements are nonzero only with the
Jordan doublet $\psi_{2,0}$ and $\psi_{2,1}$. But according
to~\eqref{delta-expansion}, $\psi_{2,1}$ is not included in the expansion of
the delta function, while $\psi_{2,0}$ is an eigenfunction and does not generate 
a linear term during evolution. As a result, we again come to
expression~\eqref{corr-lin-semiinf}.

A comparison of expressions~\eqref{corr-lin-inf} and~\eqref{corr-lin-semiinf}
shows that despite the presence of the additional factor $\omega t$
in Eq.~\eqref{corr-lin-inf}, the Lyapunov exponent can be standardly determined
from either of the correlators at large times:
\be
  \lambda
  =
  - \lim_{t\to\infty} \frac{\partial\ln\corr{\theta(0)\theta(t)}}{\partial t}
  =
  - \lim_{t\to\infty} \frac{\partial\ln\corr{\theta(t)}_{\theta_0}}{\partial t} .
\ee

\textbf{5.}
We now proceed to calculating the Lyapunov exponent for the non-FT \emph{for
arbitrary values of the parameter} $\alpha/\omega^3$. The Lyapunov exponent,
which governs the decay of the correlations at large times, is determined
by the energy of the first excited state. As shown above, in the case of
weak driving, $\lambda=\omega$. As the parameter $\alpha/\omega^3$
increases, the anharmonicity of the pendulum leads to a deviation of
$\lambda$ from $\omega$.

For a small value of the parameter $\alpha/\omega^3\ll1$, the nonlinear
terms in Eq.~\eqref{L-FP} can be taken into account perturbatively, 
which allows obtaining both a correction to the
eigenfunction $\psi_n$, which becomes dependent on the ``antiholomorphic''
coordinate $\overline{z}$, and a correction to the eigenvalue $\epsilon_n$.
This procedure looks especially simple for the first excited state, which
is nondegenerate and has no generalized eigenfunctions. For this, we represent
the eigenfunction and the corresponding energy as power series in the small
parameter $x=\alpha/\omega^3$:
\begin{gather}
\nonumber
  \psi_1 = [1 + h_1(z,\overline{z}) x + h_2(z,\overline{z}) x^2 + \dots] e^{-z^2} ,
\\
\nonumber
\epsilon_1 = \omega(1 + \gamma_1 x + \gamma_2 x^2 + \dots) ,
\end{gather}
where $h_m(z,\overline{z})$ is a polynomial of a degree not exceeding $4m$.
Substituting these expressions in the equation $L\psi_1=\epsilon_1\psi_1$
and solving sequentially in each order in $x$, we can calculate the first
few polynomials $h_m(z,\overline{z})$ and the coefficients $\gamma_m$. The
result for $\epsilon_1$ defining the Lyapunov exponent is given in
Eq.~\eqref{lambda-result}.

A similar approach allows also finding corrections to the zero
mode~\eqref{psi0-lin} of the superpotential $\psi_0$ in powers of
$\alpha/\omega^3$. As anticipated from the supersymmetry of the theory, 
its energy remains zero. The found corrections allow obtaining an
analytic expansion for the one-point statistics of the NFT, calculated
numerically in~\cite{we}. In particular, they allow refining
formula~\eqref{th2-lin} for $\corr{\theta^2}$,
\be
\label{th2-series}
  \corr{\theta^2}
  =
  \frac{x}{2}
  -\frac{13}{16} x^2
    +\frac{26989}{12288}x^3
  + \dots ,
\ee
and also describing the non-Gaussianity of the distribution function
$P(\theta)$ characterized by the fourth cumulant $\corr{\corr{\theta^4}}=
{\corr\theta^4}-3\corr{\theta^2}^2$:
\be
\label{th2-series}
  \corr{\corr{\theta^4}}
  =
	-\frac{241}{256} x^3+\frac{64725}{8192}x^4
	+\dots 
\ee
Note that the difference from the normal distribution measured by the
kurtosis $\corr{\corr{\theta^4}}/\corr{\theta^2}^2$ occurs only in the first
order in $x=\alpha/\omega^3$. A negative value of $\corr{\corr{\theta^4}}$
is related to suppression of the tails of $P(\theta)$ due to the finiteness
of the interval $(-\pi/2,\pi/2)$.

In the case of an arbitrary noise strength, the excited states of
operator~\eqref{L-FP} can be constructed only numerically. To determine the
Lyapunov exponent $\lambda=\epsilon_1$, we must find the first excited state
by solving the equation $L\psi=\epsilon_1\psi$ with the boundary conditions
\begin{subequations}
\label{BC-excited}
\begin{gather}
\psi(\pi/2,p<0)=\psi(\theta,-\infty)=0,
\\
\psi(-\pi/2,p>0)=\psi(\theta,\infty)=0.
\end{gather}
\end{subequations}
These boundary conditions are similar to the boundary conditions for the
zero mode of the superpotential derived in Ref.\ \cite{we}, with the only
difference that in the part of the boundary where the wave function is
specified, its value is zero and not $\pm1/2$.

\begin{figure}
\includegraphics[width=\columnwidth]{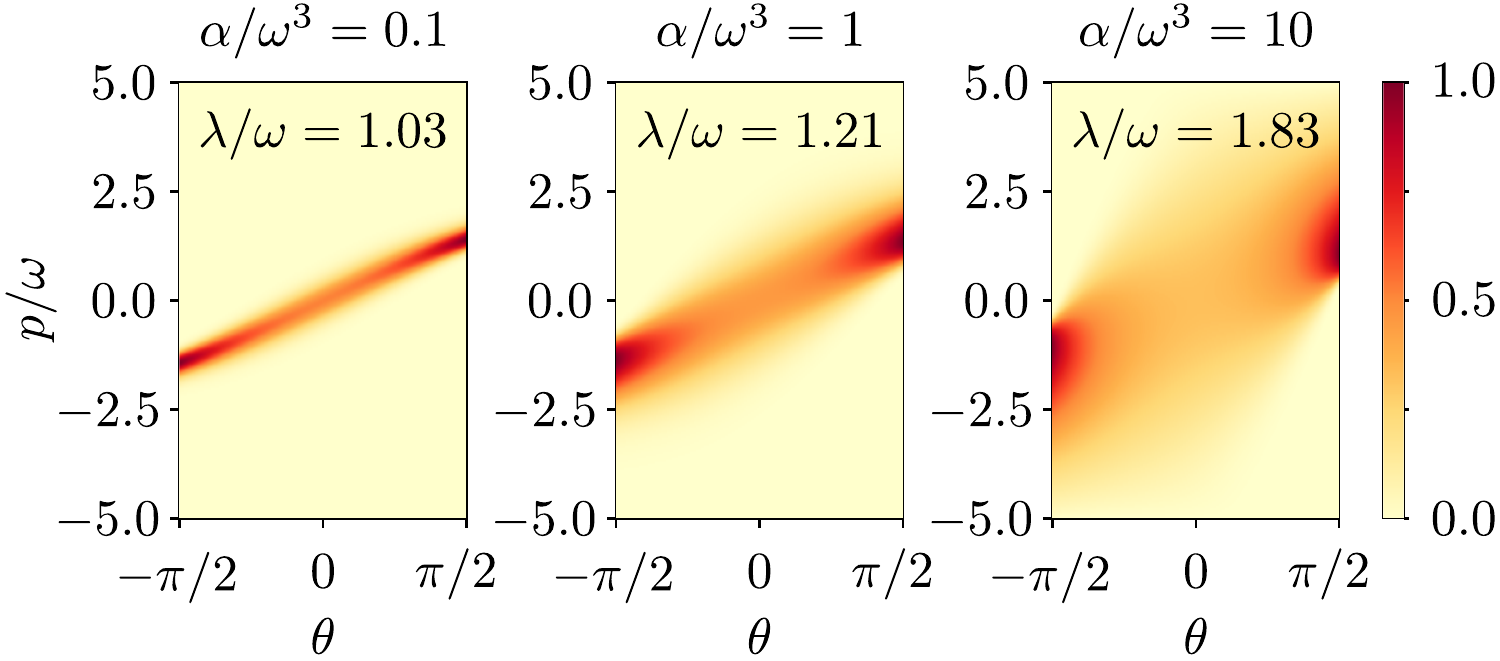}
\caption{The first excited state $\psi_1(\theta,p)$ of the operator~\eqref{L-FP}
for three values of the parameter $\alpha/\omega^3=0.1,\,1,\,10$. The wave
function is normalized to the maximum value.}
\label{F:first-mode}
\end{figure}

In Fig.~\ref{F:first-mode}, we show the first excited state 
determined numerically for various values of the parameter $\alpha/\omega^3$.
For small $\alpha/\omega^3$, the function $\psi_1(\theta,p)$ is close to the
Gaussian $\psi_{1,0}(z)$, slightly increasing near $\theta=\pm\pi/2$. As
$\alpha/\omega^3$ increases, the maximum of $\psi_1(\theta,p)$ near the
boundaries of the interval become more pronounced, and at
$\alpha/\omega^3\to\infty$, the first mode has two humps localized near the
boundaries. In Fig.~\ref{F:lambda}, we plot the energy of the first mode
(which determines the Lyapunov exponent) as a function of the parameter
$\alpha/\omega^3$. For small $\alpha/\omega^3$, the numerical calculation
agrees with expression~\eqref{lambda-result} obtained using the perturbation
theory up to the values $\alpha/\omega^3\approx0.25$. For large $a/\omega^3$, 
the Lyapunov exponent in units of $\omega$ can be expanded in
powers of $(\alpha/\omega^3)^{1/3}$ with the leading term $\lambda\approx
0.66\,\alpha^{1/3}$.

In conclusion, we note that the developed theory is a generalization
of the supersymmetric approach proposed in Ref.\ \cite{we} to the case of 
a non-FT on finite time intervals and to multipoint correlation functions.
The suggested classification of the excited states of the transfer-matrix
Hamiltonian completes the construction of the theory of the statistical
properties of a non-FT in the Whitney problem with random short-range
driving. The developed formalism allows finding any correlation functions on
a non-FT by solving partial differential equations of the Fokker--Planck
type with specific boundary conditions.

The authors thank A.~V.~Khvalyuk and I.~V.~Poboiko for the help with
numerical calculations. This work was supported by a grant from the Russian
Science Foundation (Project No.~20-12-00361).

\end{document}